**Li-O$_2$ cell scale energy densities**

Christian Prehal and Stefan A. Freunberger

In a recent issue of Joule, Dongmin Im and coworkers from Samsung in South Korea describe a prototype lithium-O$_2$ battery that reaches ~700 Wh kg$^{-1}$ and ~600 Wh L$^{-1}$ on cell level. They cut all components to the minimum to reach this value. Difficulties to fill the pores with discharge product and inhomogeneous cell utilization turn out to limit the achievable energy. The work underlines the importance to report performance with respect to full cell weight and volume.

Mobile energy sources are powering modern society. Lithium-ion batteries were key drivers for the recent portable electronics revolution but are reaching the limits of possible energy storage per unit mass.[1] With the drive to adapt them more widespread into large-scale applications like electric vehicles, properties like sustainability and cost become ever more important[2]. Research into alternative technologies is hence need to serve these additional requirements. Amongst the few options, metal-air batteries have attracted tremendous attention because they theoretically hold much higher energy per unit mass than current Li-ion batteries.[1,3,4] They use lithium metal (but also variants with sodium or potassium exist) at the anode and oxygen – drawn from air – at the cathode. In the Li-O$_2$ cell the active material at the cathode is Li$_2$O$_2$, which forms according to electrochemical reaction O$_2$ + 2 e$^-$ + 2 Li$^+$ $\leftrightarrows$ Li$_2$O$_2$. This way they avoid expensive and toxic cobalt as used in current lithium-ion batteries. However, akin to many so called "beyond-intercalation battery chemistries" reporting performance metrics that are meaningful in comparison to established battery chemistries is tricky.[5,6] Numbers based on the active material alone, as done for Li-ion materials, is prone to hugely overemphasize the gain in energy with the new battery chemistries. Works such as now reported by Im and coworkers will provide more realistic views on the achievable performance.

Li-O$_2$ batteries with exceptionally high energies have been claimed many times before. Numbers were based on extraordinarily high capacities per carbon electrode mass, which in some cases reached several 10,000 mAh·g$_C$$^{-1}$.[5] Such values compare superficially very favorably to ~100–300 mAh·g$^{-1}$ of active intercalation material in Li-ion batteries. However, to judge true electrode performance all active and inactive components needs to be included. By doing so, Im and co-workers[7] achieve a cell scale specific energy of 700 Wh·kg$_{cell}$$^{-1}$, considering the mass of all cell components of the folded cell structure. Decisive for capacity is the degree with which the initially electrolyte filled pores of the electron conducting porous electrode can be filled with Li$_2$O$_2$ (Fig. 1A, B). This can only be achieved if nanopores are filled uniformly across the cathode by a large fraction with active material $(Li_2O_2)$[5]. To achieve high specific cell energies both microscopic limitations[4,8,9] and parameters defining the uniformity on a macroscopic (cell) scale need to be identified.

Given its central importance for capacity and rate capability, the way how $Li_2O_2$ forms in the porous cathode has been subject to intensive investigations over more than a decade.[3,8,9] It is now widely accepted that $Li_2O_2$ may form via either a surface or solution mechanism, which results in either conformal coating of the pore surface (Fig. 1D) or in large, typically toroidal $Li_2O_2$ particles (Fig. 1C). Surface coating is limited by the layer thickness that still allows electron tunnelling, which has been estimated to be ~10 nm. The solution mechanism allows for much larger capacities since diffusion of intermediates provides better mobility than tunnelling provides for electrons.

Simulations of the authors[7] show that pore filling with $Li_2O_2$ is heterogeneous along the carbon cathode in the direction where $O_2$ is transported through a gas diffusion layer (Fig. 1A, B), even in their optimized folded cell architecture. The simulation considers the capacity limitation by the increasing ohmic overpotential of the passivating, epitaxial layer of $Li_2O_2$ (following the surface mechanism, Fig. 1D). This is a limitation of their work since in reality both surface and solution mechanism may prevail. Whenever $O_2$, $Li^+$ or electron transport is limited on a macroscopic level, $Li_2O_2$ will be non-uniformly distributed, both in terms of pore filling and morphology. Thus, (global) transport on the cell level has always a significant impact on (local) areal current densities $(mA \cdot cm^{-2})$. Since the reaction mechanism on a microscopic scale (Fig. 1A vs. Fig. 1B) is known to be strongly dependent on the applied current densities[4,8-10], the $Li_2O_2$ morphologies will be as well. In turn $Li_2O_2$ morphologies define the maximum achievable pore filling and capacities. Consequently, macroscopic engineering and fundamental microscopic aspects should not be considered independently from each other. This applies both for interpreting sophisticated (operando) experimental data obtained in complicated operando cell designs or performance metrics obtained with standard cell assemblies that are used for battery characterization.

Despite a body of work already published on understanding the morphology evolution, the picture is not yet complete. Specifically, there is no informed choice on how to pick the best pore size distribution for given electrolyte properties. To conclusively understand capacity limitation in $Li-O_2$ batteries, we need structure-sensitive methods that allow correlating the $Li_2O_2$ structure on a microscopic and a macroscopic cell level with the integral cell performance.

*Christian Prehal and Stefan A. Freunberger\**
*Graz University of Technology, Stremayrgasse 9, 8010 Graz, Austria.*

*e-mail: freunberger@tugraz.at

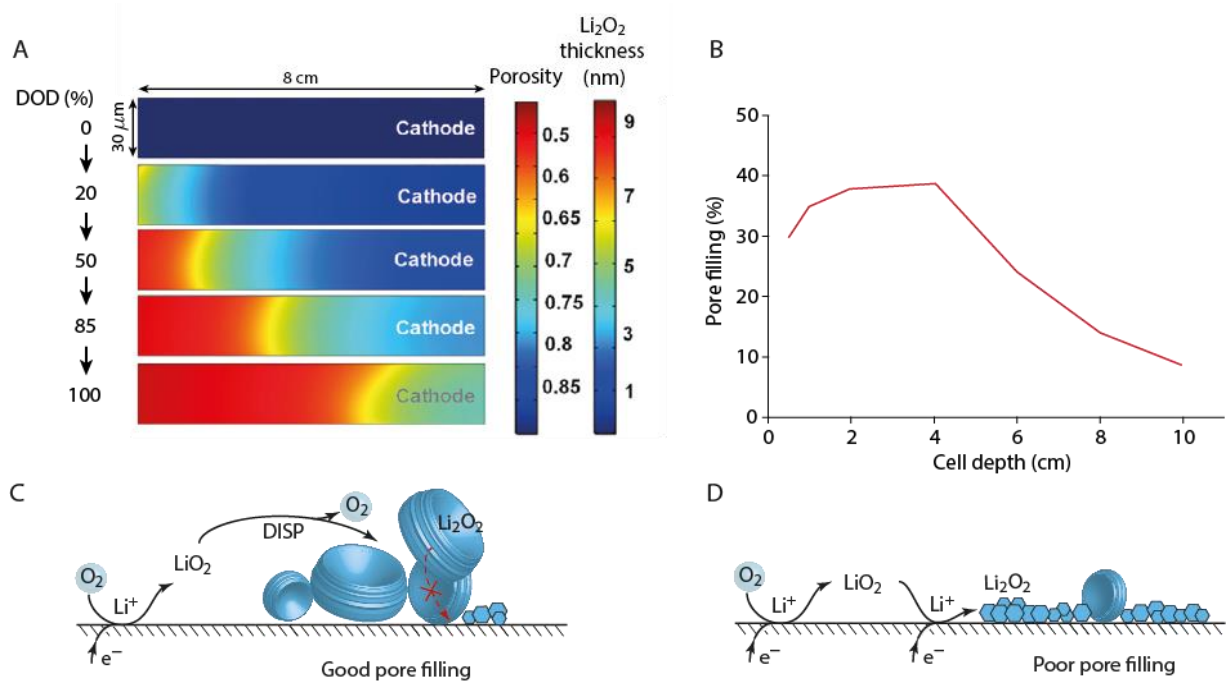

**Figure 1. Heterogeneity of Li$_2$O$_2$ pore filling in a Li-O$_2$ battery cathode.** (A), Remaining porosity and Li$_2$O$_2$ thickness for different depths of discharge (DOD) across a 30 µm thick and 8 cm long cathode, resulting from simulations in Im et al.[7]. (B), Li$_2$O$_2$ pore filling across a 20 µm thick Li-O$_2$ battery cathode, calculated from energy densities presented in Im et al[7]. (C), Good pore filling on a local scale is achieved by promoting the formation of large toroidal particles via a disproportionation reaction of LiO$_2$ (solution mechanism). (D), Poor utilization of the available pore space on a local scale by forming a conformal coating of Li$_2$O$_2$ via a surface mechanism. Fig. 1C and D are adopted with modifications from Ref. 4 - Published by The Royal Society of Chemistry.